\documentclass[]{aipproc}

\layoutstyle{6x9}

\SetInternalRegister\hbadness{8000} 

\begin{document}
\begin{flushright}
MPI-PHT-2001-38\\
September 2001
\end{flushright}
\title{$B\to V\gamma$ at NLO from QCD Factorization\footnote{
Invited talks given at the Workshop on Exclusive B Decays, 20--22 July 2001, Regensburg and at the 9th International Symposium on Heavy Flavor Physics, 10--13 September 2001, Caltech, Pasadena.
Based on work together with Gerhard Buchalla \cite{BB}}
}


\author{Stefan W. Bosch}{
  address={Max-Planck-Institut f\"ur Physik,
F\"ohringer Ring 6, D-80805 Munich, Germany},
  email={bosch@mppmu.mpg.de}
}

%

\copyrightyear  {2001}

\begin{abstract}
We discuss the exclusive radiative $B$-meson decays $B\to K^*\gamma$ and $B\to\rho\gamma$ in a model-independent manner. The analysis is based on the heavy-quark limit of QCD. This allows a factorization of perturbatively calculable contributions to the $B\to V\gamma$ matrix elements from non-perturbative form factors and universal light-cone distribution amplitudes. These results allow us to compute {\it exclusive\/} $b\to s(d) \gamma$ decays systematically beyond the leading logarithmic approximation. We present results for these decays complete to next-to-leading order in QCD and to leading order in the heavy-quark limit. Phenomenological implications for various observables of interest are discussed, including direct CP violation and isospin breaking effects.
\end{abstract}

\date{\today}

\maketitle

\section{Introduction}

The radiative transitions $b\to s(d)\gamma$ are among the most valuable probes of flavour physics. Although they are rare decays the Cabibbo-favoured $b\to s\gamma$ modes are experimentally accessible already at present. The inclusive branching fraction has been measured to be
\begin{equation}\label{bsgamex}
B(B\to X_s\gamma)=(3.23\pm 0.42)\cdot 10^{-4}
\end{equation}
combining the results of \cite{CHENHON,BAR,ABENAK}.
The branching ratios for the exclusive channels have been determined by CLEO \cite{COA}, and more recently also by BaBar \cite{SPARYD} and BELLE \cite{NAK} to give the averaged values:
\begin{eqnarray}\label{b0kgamex}
B(B^0\to K^{*0}\gamma)=(4.51\pm 0.54)\cdot 10^{-5}\\
\label{bpkgamex}B(B^+\to K^{*+}\gamma)=(3.86\pm 0.59)\cdot 10^{-5}.
\end{eqnarray}
On the theoretical side, the flavour-changing neutral current (FCNC)
reactions $b\to s(d)\gamma$ are characterized by their high sensitivity
to New Physics and by the particularly large impact of short-distance
QCD corrections \cite{BBM,AGM,BMMP,CFMRS}.
Considerable efforts have therefore been devoted to achieve a full
calculation of the inclusive decay $b\to s\gamma$ at next-to-leading
order (NLO) in renormalization group (RG) improved perturbation
theory \cite{AY,GHW,CMM} (see \cite{MG} for recent reviews).

Whereas the inclusive mode can be computed perturbatively, using
the fact that the $b$-quark mass is large and employing the heavy-quark
expansion, the treatment of the exclusive channel $B\to K^*\gamma$
is in general more complicated. In this case bound state effects
are essential and need to be described by nonperturbative 
hadronic quantities (form factors).
The basic mechanisms at next-to-leading order were already
discussed previously for the $B\to V\gamma$ amplitudes \cite{AAWGSW}.
However, hadronic models were used to evaluate the various
contributions, which did not allow a clear separation
of short- and long-distance dynamics and a clean distinction
of model-dependent and model-independent features.

In this talk we present the results of \cite{BB} where a systematic analysis of the exclusive radiative decays $B\to V\gamma$ ($V=K^*$, $\rho$) in QCD, based
on the heavy quark limit $m_b\gg\Lambda_{QCD}$, was performed. We quote factorization formulas for the evaluation of the relevant hadronic matrix elements of local operators in the weak Hamiltonian. A similar subject was treated in \cite{BFS,AP}.
Factorization holds in QCD to leading power in the heavy quark limit.
This result relies on arguments similar to those used previously
to demonstrate QCD factorization for hadronic two-body modes
of the type $B\to\pi\pi$ \cite{BBNS1,BBNS2}. 

This framework allows us to separate perturbatively calculable
contributions from the nonperturbative form factors and universal meson
light-cone distribution amplitudes (LCDA) in a systematic way. 
This includes the treatment of loop effects from light quarks,
in particular up and charm.
Such loop effects are straightforwardly included for the inclusive
decays $b\to s(d)\gamma$. For the exclusive modes, however,
the effects from virtual charm and up quarks have so far been
considered to be uncalculable ``long-distance'' contributions
and have never been treated in a model independent fashion.

Finally, power counting in $\Lambda_{QCD}/m_b$  implies a hierarchy
among the possible mechanisms for $B\to V\gamma$ transitions. This
allows us to identify leading and subleading contributions.
For example, weak annihilation contributes only at subleading
power in the heavy quark limit.

Within this approach, higher order QCD corrections can be consistently
taken into account. We give the $B\to V\gamma$ decay amplitudes
at next-to-leading order (NLO). Furthermore numerical values for CP-asymmetries and isospin-violating ratios are given. After including NLO corrections the largest uncertainties still come from the $B\to V$ form factors, which are at present known only with limited precision ($\sim \pm 15\%$), mostly from QCD sum rule calculations \cite{BB98}. The situation should improve in the future with the help of both lattice QCD \cite{FS} and analytical methods based on the heavy-quark and large-energy limits \cite{CLOPR,BF,BH}.



\section{Basic Formulas}

The effective Hamiltonian for $b\to s\gamma$ transitions reads
\begin{equation}\label{heff}
{\cal H}_{eff}=\frac{G_F}{\sqrt{2}}\sum_{p=u,c}\lambda_p^{(s)}
\left[ C_1 Q^p_1 + C_2 Q^p_2 +\sum_{i=3,\ldots ,8} C_i Q_i\right]
\end{equation}
where $\lambda_p^{(s)}=V^*_{ps}V_{pb}$. The relevant operators are given by
\begin{eqnarray}\label{q1def}
Q^p_1 &=& (\bar sp)_{V-A}(\bar pb)_{V-A} \\
Q^p_2 &=& (\bar s_i p_j)_{V-A}(\bar p_j b_i)_{V-A} \\
\label{q7def}
Q_7 &=& \frac{e}{8\pi^2}m_b\, 
        \bar s_i\sigma^{\mu\nu}(1+\gamma_5)b_i\, F_{\mu\nu}\\
\label{q8def}
Q_8 &=& \frac{g}{8\pi^2}m_b\, 
        \bar s_i\sigma^{\mu\nu}(1+\gamma_5)T^a_{ij} b_j\, G^a_{\mu\nu}
\end{eqnarray}
Note that the numbering of $Q^p_{1,2}$ is reversed with respect to the convention of \cite{BBL}. We neglect the contribution from the QCD penguin operators $Q_{3\ldots 6}$, which enter at ${\cal O}(\alpha_s)$ and are further suppressed by very small Wilson coefficients. The effective Hamiltonian for $b\to d\gamma$ is obtained from
(\ref{heff}--\ref{q8def}) by the replacement $s\to d$. 

The most difficult step in computing the decay amplitudes is the
evaluation of the hadronic matrix elements of the operators in 
(\ref{heff}). A systematic treatment can be given in the heavy-quark
limit. In this case the following factorization formula is valid 
\begin{equation}\label{fform}
\langle V\gamma(\epsilon)|Q_i|\bar B\rangle =
\left[ F^{B\to V}(0)\, T^I_{i} +
\int^1_0 d\xi\, dv\, T^{II}_i(\xi,v)\, \Phi_B(\xi)\, \Phi_V(v)\right]
\cdot\epsilon
\end{equation}
where $\epsilon$ is the photon polarization 4-vector.
Here $F^{B\to V}$ is a $B\to V$ transition form factor,
and $\Phi_B$, $\Phi_V$ are leading twist light-cone distribution amplitudes
of the $B$ meson and the vector meson $V$, respectively.
These quantities are universal, nonperturbative objects. They
describe the long-distance dynamics of the matrix elements, which
is factorized from the perturbative, short-distance interactions
expressed in the hard-scattering kernels $T^I_{i}$ and $T^{II}_i$.
The QCD factorization formula (\ref{fform}) holds up to
corrections of relative order $\Lambda_{QCD}/m_b$.

\begin{figure}[t]
   \includegraphics[height=2.5cm]{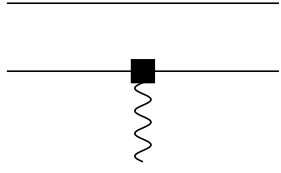}
\caption{Contribution of the magnetic penguin operator $Q_7$
described by $B\to V$ form factors. All possible gluon exchanges 
between the quark lines are included in the form factors and have not
been drawn explicitly. \label{fig:q7}}
\end{figure}

In the leading logarithmic approximation
(LO) and to leading power in the heavy-quark limit, $Q_7$ gives the only contribution to the amplitude of $\bar B\to V\gamma$ and the factorization formula (\ref{fform}) is trivial.
The matrix element is simply expressed in terms of the standard
form factor, $T^I_{7}$ is a purely kinematical function and
the spectator term $T^{II}_7$ is absent. An illustration is given
in Fig. \ref{fig:q7}.
The matrix element reads
\begin{equation}\label{q7f1f2}
\langle V(k,\eta) \gamma(q,\epsilon)|Q_7|\bar B\rangle =
-\frac{e}{2\pi^2}m_b\, c_V F_V \left[ 
\varepsilon^{\mu\nu\lambda\rho}\epsilon_\mu\eta_\nu k_\lambda q_\rho +
i (\epsilon\cdot\eta\, k\cdot q-\epsilon\cdot k\, \eta\cdot q)\right]
\end{equation}
where $c_V=1$ for $V=K^*$, $\rho^-$ and $c_V=1/\sqrt{2}$ for $V=\rho^0$.
The $\bar B\to V$ form factor $F_V$ is evaluated at momentum transfer
$q^2=0$. Our phase conventions coincide with those of \cite{BB98,ABS}.

\begin{figure}[t]
   \includegraphics[height=3cm]{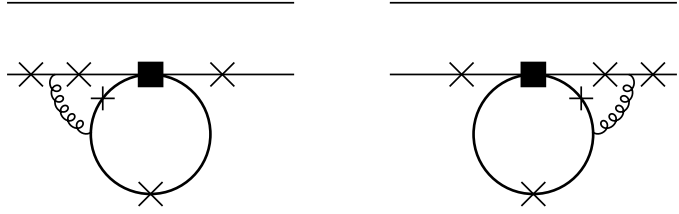}
\caption{${\cal O}(\alpha_s)$ contribution to the hard-scattering kernels $T^I_1$ from four-quark operators $Q_1$. The crosses indicate the places where the emitted photon can be attached. \label{fig:qit1}}
\end{figure}

\begin{figure}[t]
   \includegraphics[height=1.875cm]{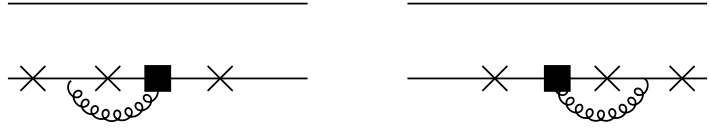}
\caption{${\cal O}(\alpha_s)$ contribution to the hard-scattering 
kernels $T^I_{8}$ from chromomagnetic penguin operator $Q_8$.
\label{fig:q8t1}}
\end{figure}

The matrix elements of $Q_1$ and $Q_8$ start contributing at ${\cal O}(\alpha_s)$. In this case the factorization formula becomes nontrivial. The diagrams for the hard-scattering kernels $T^I_{i}$ are shown in Fig. \ref{fig:qit1}
for $Q_1$ and in Fig. \ref{fig:q8t1}
for $Q_8$. These diagrams were computed in \cite{GHW} to get the virtual corrections to the inclusive matrix elements of $Q_1$ and $Q_8$. In our case they determine the kernels $T^I_{1}$ and $T^I_{8}$.
As required for the consistency of the factorization formula these corrections must be dominated by hard scales of order $m_b$ and hence must be infrared finite. This is indeed the case. Re-interpreted as the perturbative hard-scattering kernels for the exclusive process, the results from \cite{GHW} imply
\begin{equation}\label{q1me1}
\langle Q_{1,8}\rangle^I=\langle Q_7\rangle 
\frac{\alpha_s C_F}{4\pi} G_{1,8}
\end{equation}
where $C_F=(N^2-1)/(2N)$, with $N=3$ the number of colours, and
\begin{eqnarray}\label{G1}
G_1(s_c) &=& -\frac{104}{27}\ln\frac{\mu}{m_b}+ g_1(s_c) \\
\label{G8}
G_8 &=& \frac{8}{3}\ln\frac{\mu}{m_b} + g_8
\end{eqnarray}
The finite part $g_1(s_c)$ of the $Q_1$-contribution depends via $s_c=\frac{m^2_c}{m^2_b}$ on the mass of the quark running in the loop in Fig. \ref{fig:qit1}.

We now turn to the mechanism where the spectator participates
in the hard scattering.

\begin{figure}[t]
   \includegraphics[height=3cm]{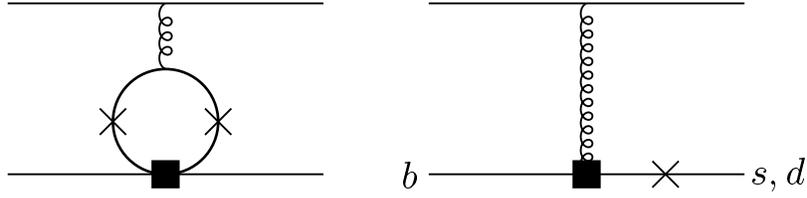}
\caption{${\cal O}(\alpha_s)$ and leading power contribution to the hard-scattering kernels $T^{II}_i$ from four-quark operators $Q_i$ (left) and from $Q_8$. \label{fig:qit2}}
\end{figure}

\begin{figure}[t]
   \includegraphics[height=6cm]{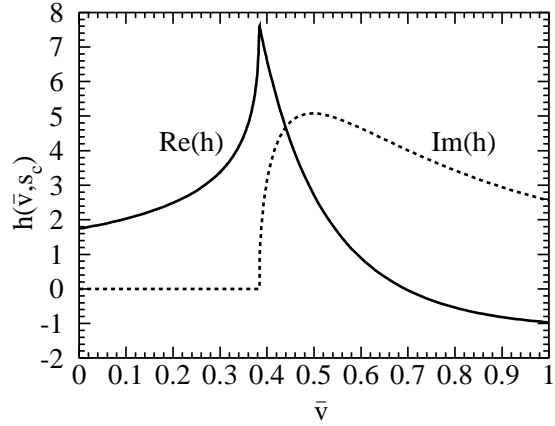}
\caption{The hard-scattering kernel $h(\bar v,s_c)$ as a function
of $\bar v$. \label{fig:hcvbar}}
\end{figure}

The non-vanishing contributions to
$T^{II}_i$ are shown in Fig.~\ref{fig:qit2}. 
To find the correction for $\langle Q_1\rangle$ we compute the first diagram in Fig. \ref{fig:qit2}
We obtain
\begin{equation}\label{q1me2}
\langle Q_1\rangle^{II}=\langle Q_7\rangle 
\frac{\alpha_s(\mu_h) C_F}{4\pi} H_1(s_c)
\end{equation}
with
\begin{equation}\label{h1s}
H_1(s)=-\frac{2\pi^2}{3 N}\frac{f_B f^\perp_V}{F_V m^2_B}
\int^1_0 d\xi\frac{\Phi_{B1}(\xi)}{\xi}\int^1_0 dv\, h(\bar v,s)
\Phi_\perp(v)
\end{equation}
The first negative moment of the light-cone distribution amplitude $\Phi_{B1}(\xi)$ can be pa\-ra\-me\-trized by $m_B/\lambda_B$ where $\lambda_B={\cal O}(\Lambda_{QCD})$. The hard-scattering kernel $h(\bar v,s_c)$ is displayed in Fig. \ref{fig:hcvbar}. It is real for $\bar v\leq 4s$ and develops an imaginary part for the light-cone momentum fraction of the anti-quark in the vector meson $\bar v > 4s$.

The correction to $\langle Q_8\rangle$ from the hard spectator
interaction comes from the second diagram in Fig. \ref{fig:qit2}.
One finds 
\begin{equation}\label{q8me2}
\langle Q_8\rangle^{II}=\langle Q_7\rangle 
\frac{\alpha_s(\mu_h) C_F}{4\pi} H_8
\end{equation}
where
\begin{equation}\label{h8}
H_8=\frac{4\pi^2}{3 N}\frac{f_B f^\perp_V}{F_V m^2_B}
\int^1_0 d\xi\frac{\Phi_{B1}(\xi)}{\xi}
\int^1_0 dv\frac{\Phi_\perp(v)}{v}
\end{equation}
Here the LCDA $\Phi_\perp(v)$ is of leading power for a transversly polarized vector meson. The last convolution integral in (\ref{h8}) can be performed explicitely and leads to a combination of Gegenbauer moments of $\Phi_\perp$ \cite{BB96,BB98}.

There are further mechanisms that can in principle contribute
to $\bar B\to V\gamma$ decays.
One possibility is weak annihilation.
In this case the leading-power projection onto the meson $V$ vanishes because the trace over an odd number of Dirac matrices is zero. A non-vanishing result arises from the subleading-power projections onto $\Phi_\parallel$ and $g_\perp$.

Despite its power suppression, the dominant annihilation amplitude
can be computed within QCD factorization. This is because the
colour-transparency argument applies to the emitted, highly energetic
vector meson in the heavy-quark limit \cite{BBNS2,GP}.

Since weak annihilation is a power correction, we will content 
ourselves with the lowest order result (${\cal O}(\alpha^0_s)$)
for our estimates below. In particular, we shall include the 
annihilation effects from operators $Q_{1,2}$ to estimate
isospin-breaking corrections in $B\to\rho\gamma$ decays.
The reason for including this class of power corrections is that they
come with a numerical enhancement from the large Wilson
coefficients $C_{1,2}$ ($C_1\approx 3|C_7|$) and are not CKM suppressed.
Instead, a CKM suppression of annihilation effects occurs for
$B\to K^*\gamma$ and these contributions are thus very small in this case.

\section{Results}
Finally, we can combine these results and write, adding the up- and charm- quark contribution,
\begin{equation}\label{abvgam}
A(\bar B\to V\gamma)=\frac{G_F}{\sqrt{2}}
\left[\sum_{p=u,c} \lambda_p^{(s)}\, a^p_7(V\gamma)\right] 
\langle V\gamma|Q_7|\bar B\rangle
\end{equation}
where, at NLO, the factorization coefficients $a^p_7(V\gamma)$ are given as
\begin{eqnarray}\label{a7vgam}
a^p_7(V\gamma) = C_7 &+& \frac{\alpha_s(\mu) C_F}{4\pi}
\left( C_1(\mu) G_1(s_p)+ C_8(\mu) G_8\right) \nonumber \\  
  &+& \frac{\alpha_s(\mu_h) C_F}{4\pi} 
  \left( C_1(\mu_h) H_1(s_p)+ C_8(\mu_h) H_8\right)
\end{eqnarray}
Here the NLO expression for $C_7$ \cite{CMM} has to be used, while the leading
order values are sufficient for $C_1$ and $C_8$. 
Numerically we obtain for central values of all input parameters, at $\mu=m_b$, and displaying separately the size of the various correction terms:
\begin{eqnarray}\label{numa7cK}
a_7^c(K^* \gamma) &=& \begin{array}[t]{cccc}
-0.3221 & +0.0113 & -0.0820-0.0147i & -0.0144-0.0109i\\
C_7^{LO} & \Delta C_7^{NLO} & T^I_{1,8}\mbox{-contribution} & 
  T^{II}_{1,8}\mbox{-contribution}\\
\end{array} \nonumber\\
&=& -0.4072 -0.0256i.
\end{eqnarray}
We note a sizable enhancement of the leading order value, dominated by the 
$T^I$-type correction.
A complex phase is generated at NLO, 
where the $T^I$-corrections and the hard-spectator interactions ($T^{II}$) 
yield comparable effects.

\begin{figure}[t]
\includegraphics[height=8cm]{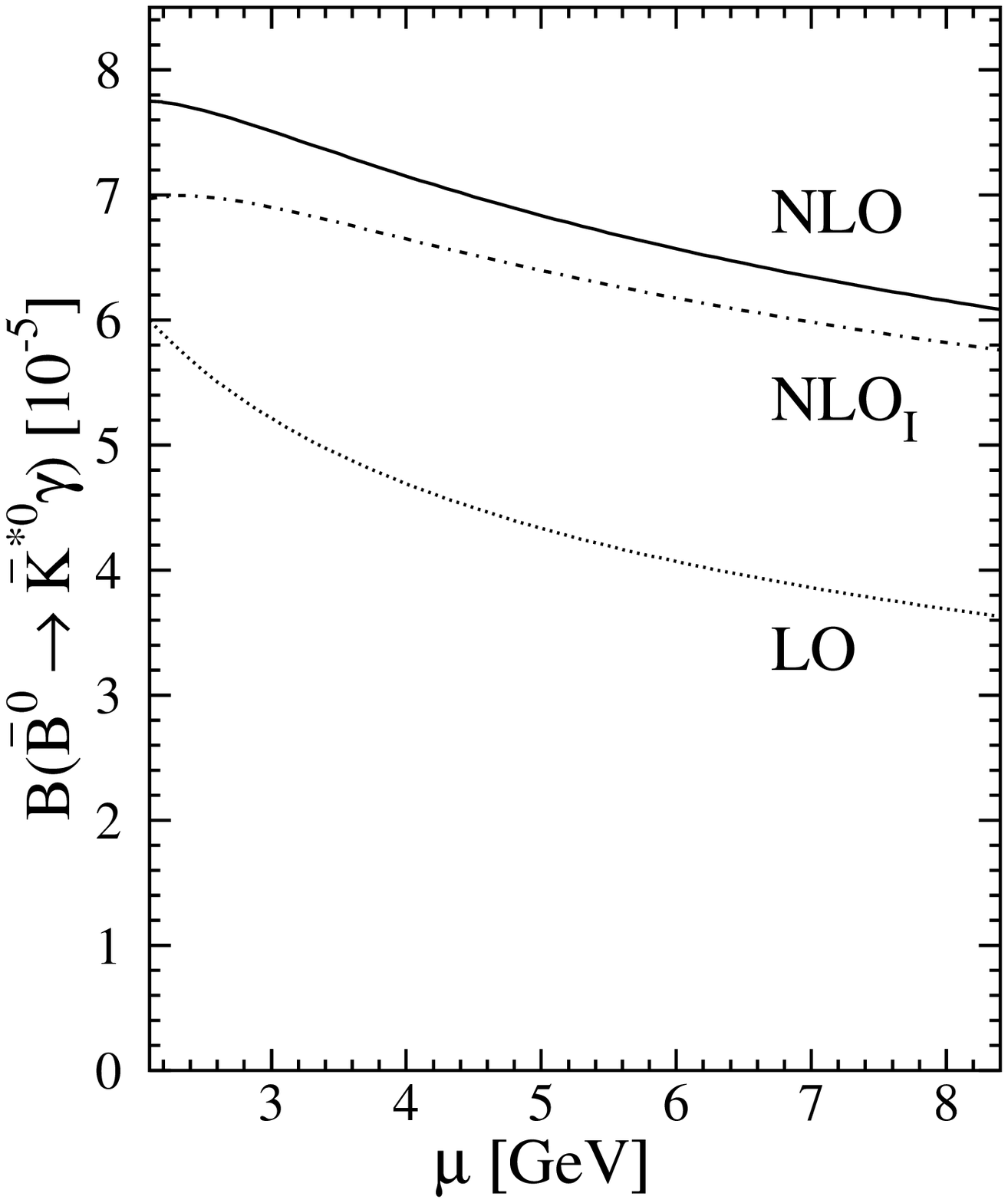}\includegraphics[height=8cm]{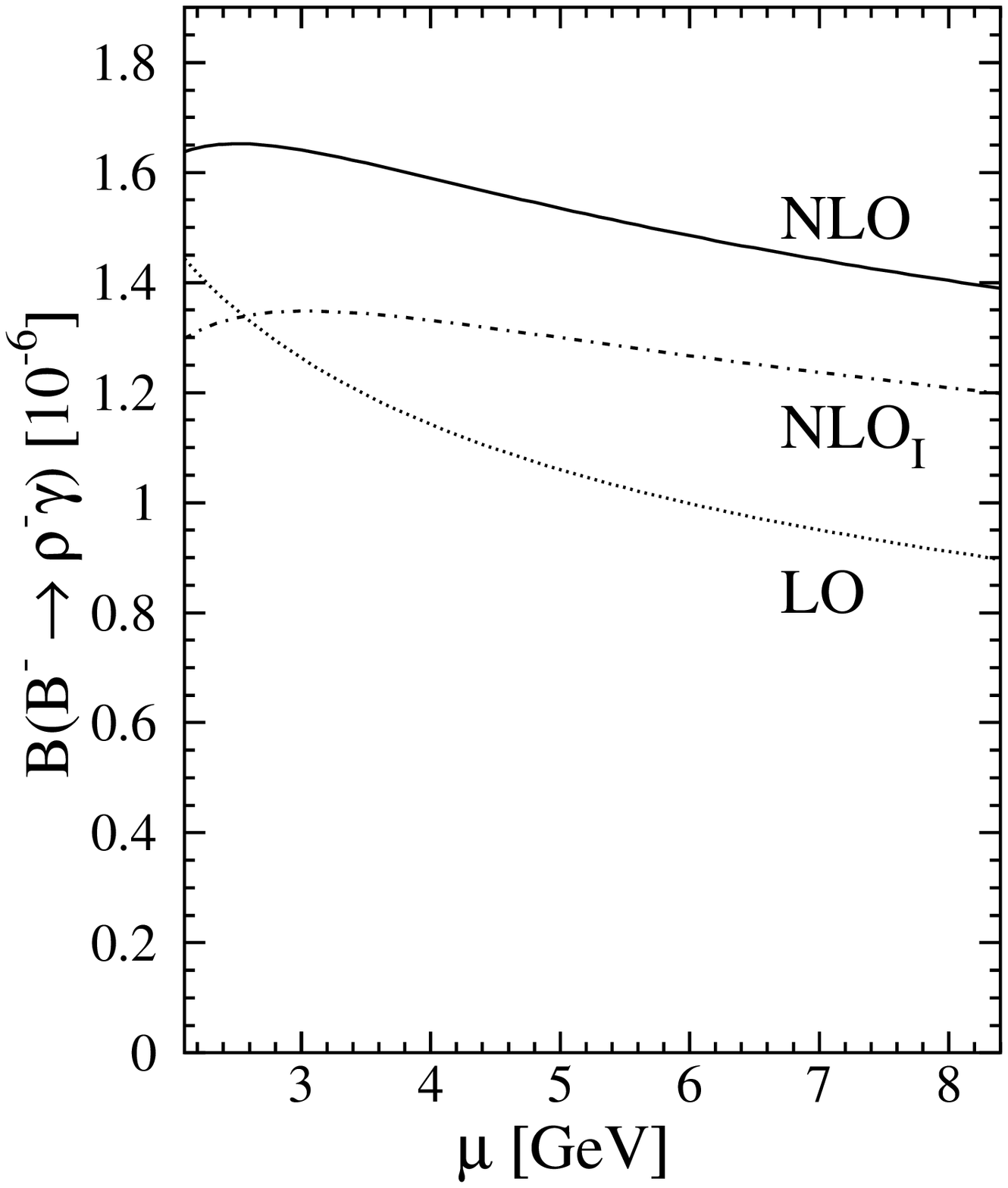}
\caption{Dependence of the branching fractions $B(\bar{B}^0 \to \bar{K}^{*0} \gamma)$ and $B(B^- \to \rho^- \gamma)$ on the renormalization scale $\mu$. The dotted line shows the LO, the dash-dotted line the NLO result including type-I corrections only and the solid line shows the complete NLO result.}
\label{fig:bkrhomu}
\end{figure}

The net enhancement of $a_7$ at NLO leads to a corresponding enhancement 
of the branching ratios, for fixed value of the form factor. This is 
illustrated in Fig. \ref{fig:bkrhomu}, where we show the residual scale 
dependence for $B(\bar{B}\to \bar{K}^{*0}\gamma)$ and $B(B^-\to\rho^-\gamma)$ 
at leading and next-to-leading order.
As expected, the inclusion of the hard-vertex corrections ($T^I$) reduces the scale dependence coming from the Wilson coefficients. The scale dependence of the complete NLO result is ``deteriorated'' because the type-II contributions appear at ${\cal O}(\alpha_s)$ for the first time and therefore introduce a completely new scale dependence.

%

%
For the decay $\bar B\to\rho\gamma$ both sectors of the effective 
Hamiltonian have the same order of magnitude.
The amplitude for the CP-conjugated mode $B\to\rho\gamma$ 
is obtained by replacing $\lambda_p^{(d)} \to \lambda_p^{(d)*}$. 
We may then consider the CP asymmetry
\begin{equation}\label{acpbrgdef}
{\cal A}_{CP}(\rho\gamma)=
\frac{\Gamma(B\to\rho\gamma)-\Gamma(\bar B\to\rho\gamma)}{
      \Gamma(B\to\rho\gamma)+\Gamma(\bar B\to\rho\gamma)}
\end{equation}

It is substantial for the $\rho\gamma$ modes and much less dependent on the form factors. Here the largest theoretical uncertainty comes from the scale dependence. This is to be expected because the direct CP asymmetry is proportional to the perturbative strong phase difference, which arises at ${\cal O}(\alpha_s)$ for the first time. Unknown power corrections could have some impact on the prediction.

A further interesting observable is the charge averaged isospin breaking ratio
\begin{equation}
\label{isobreak}
\Delta(\rho\gamma)=\frac{\Gamma(B^+\to\rho^+\gamma)}{4\Gamma(B^0\to\rho^0\gamma)}+\frac{\Gamma(B^-\to\rho^-\gamma)}{4\Gamma(\bar{B}^0\to\rho^0\gamma)}-1
\end{equation}
Within our approximations, isospin breaking is generated by weak annihilation. Isospin breaking was already discussed in \cite{AHL}, partially including NLO corrections.

\begin{figure}[t]
   \includegraphics[height=6cm]{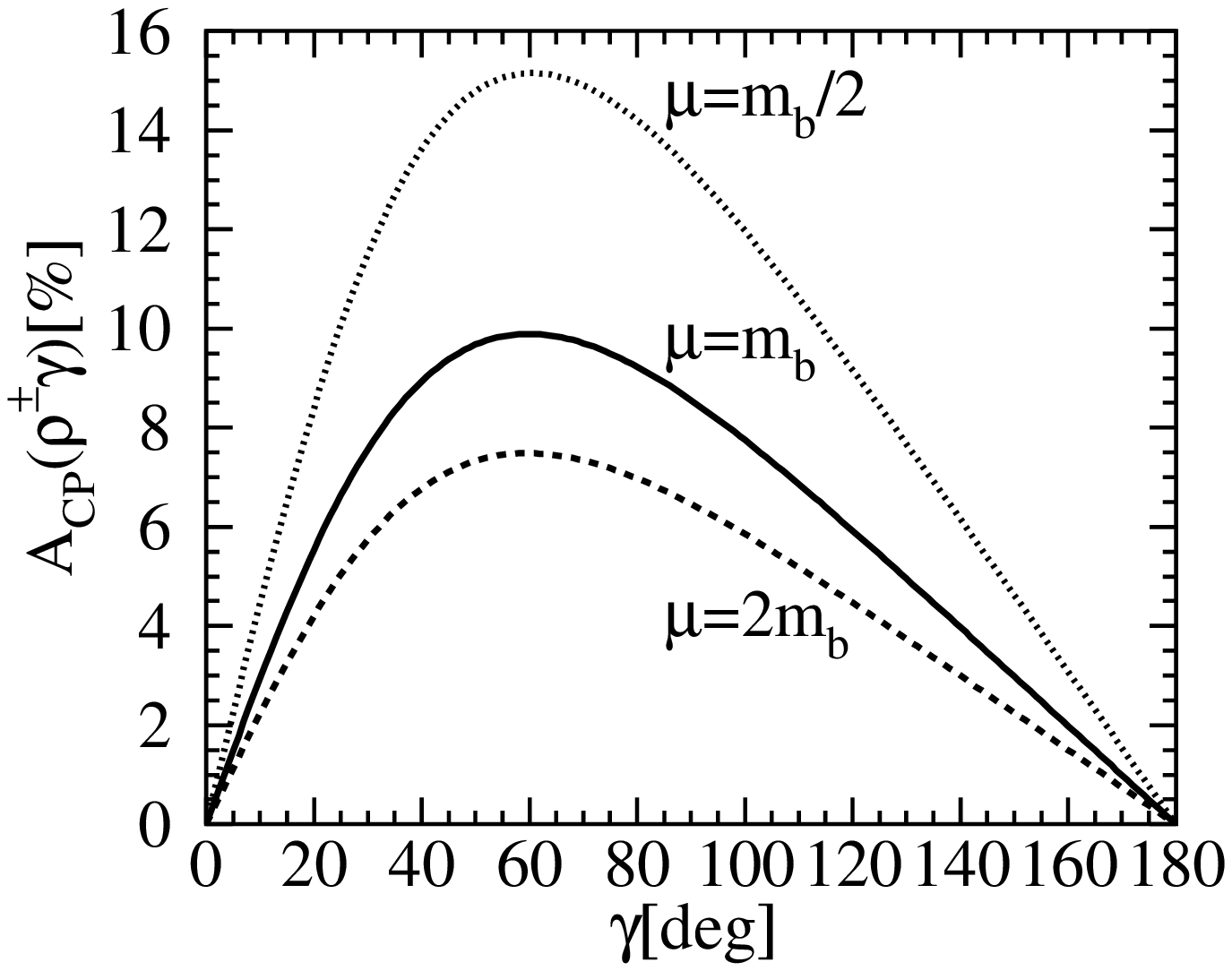}\includegraphics[height=6cm]{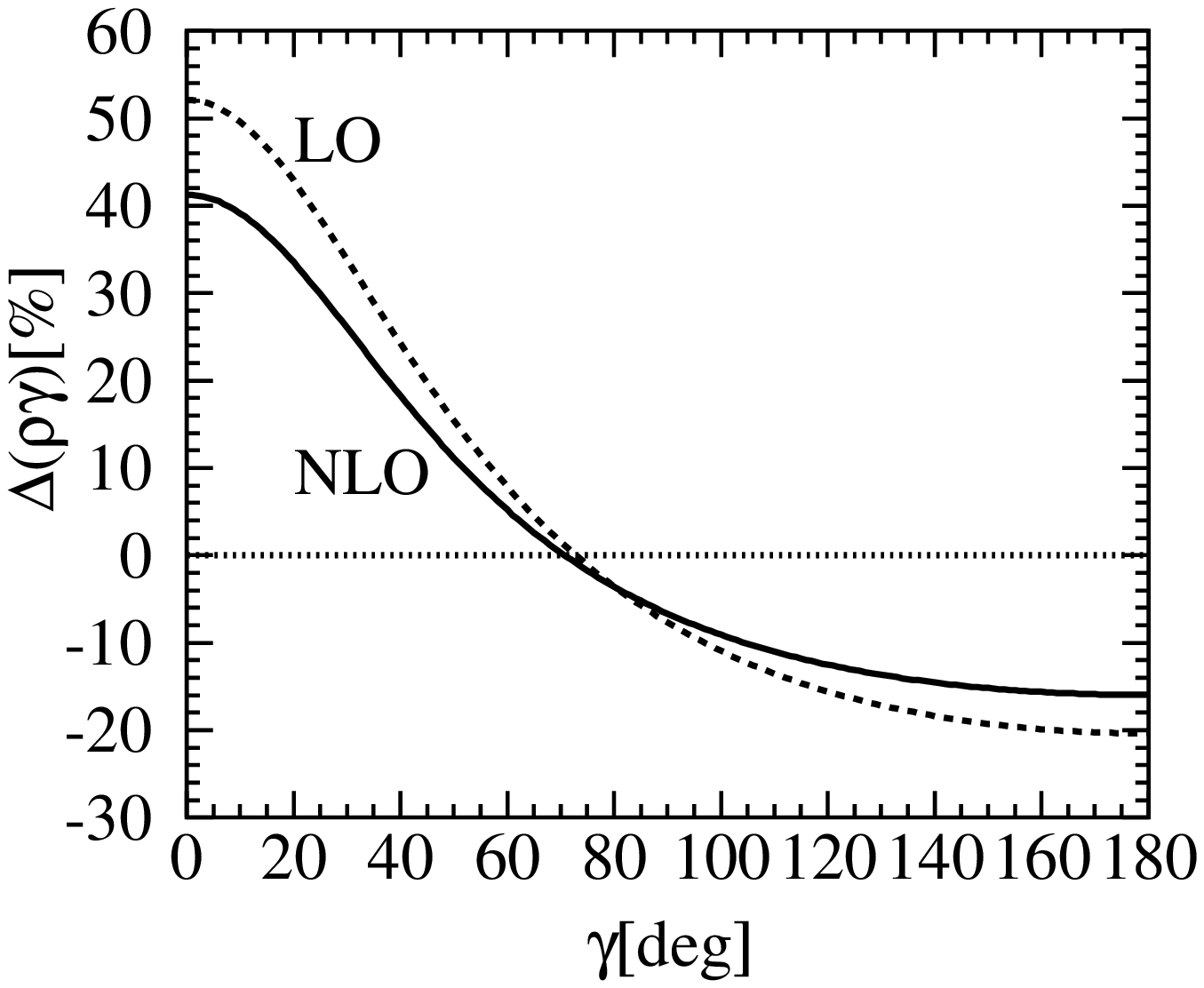}
\caption{Left: The CP asymmetry ${\cal A}_{CP}(\rho\gamma)$ as a function of the CKM angle $\gamma$ for three values of the renormalization scale $\mu=m_b/2$, $m_b$ and $2 m_b$. Right: The isospin-breaking asymmetry $\Delta(\rho\gamma)$ as a function of the CKM angle $\gamma$ at leading and next-to-leading order.\label{fig:acpiso}}
\end{figure}

$B\to\rho\gamma$ also depends sensitively on fundamental CKM parameters, such as $\left|V_{ub}/V_{cb}\right|$ and $\gamma$, and can thus in principle serve to constrain the latter quantities once measurements become available. This is further illustrated in Fig. \ref{fig:acpiso}, where the dependence on $\gamma$ is shown for ${\cal A}_{CP}(\rho\gamma)$ and $\Delta(\rho\gamma)$, respectively.
We remark that our sign of $\Delta(\rho\gamma)$ differs from the one found in \cite{AHL}.

Another application of our results concerns an estimate of U-spin breaking effects in $B\to V\gamma$ decays \cite{GRG,HM,RF}. Let us state here only that U-spin breaking effects can be sizeable. For further details we refer to \cite{BB}.

\section{Conclusions}

In this talk we have presented a systematic and
model-independent framework for the exclusive radiative
decays $B\to V\gamma$ based on the heavy-quark limit.
This allowed us to compute the decay amplitudes for these modes
consistently at next-to-leading order in QCD.

An important conceptual aspect of this analysis is the interpretation
of loop contributions with charm and up quarks, which come from
leading operators in the effective weak Hamiltonian.
We have argued that these effects are calculable in terms of
perturbative hard-scattering functions and universal meson
light-cone distribution amplitudes. They are ${\cal O}(\alpha_s)$
corrections, but are leading power contributions in the
framework of QCD factorization. This picture is in contrast to the
common notion that considers charm and up-quark loop effects as
generic, uncalculable long-distance contributions.
Non-factorizable long-distance corrections may still exist, but
they are power-suppressed.

Another important feature of the NLO calculation are the strong interaction phases, which are calculable at leading power. They play a crucial role for CP violating observables. We have seen that weak-annihilation amplitudes are power-suppressed, but can be numerically important for $B\to\rho\gamma$ because they enter with large coefficients. These effects also turn out to be calculable and were included in our phenomenological discussion at leading order in QCD.

Finally, a numerical analysis leads to our predictions for the branching ratios $B(\bar B^0\to\bar K^{*0}\gamma)\sim 7.09\cdot 10^{-5}$ and $B(B^-\to\rho^-\gamma)\sim 1.58\cdot 10^{-6}$, and for the CP asymmetries ${\cal A}_{CP}(K^*\gamma)\sim -0.5\%$ and ${\cal A}_{CP}(\rho\gamma)\sim 10\%$ for central values of the input parameters. The uncertainties for the branching ratios are of ${\cal O}(30\ldots 35\%)$ where the form factors $F_{K^*}$ and $F_\rho$ taken from \cite{BB98} clearly dominate this uncertainty. This situation however can be systematically improved. In particular, our approach allows for a consistent perturbative matching of the nonperturbative form factor to the short-distance part of the amplitude.
Power corrections are not expected to be unreasonably large because with the annihilation contribution we have included the probably numerically largest of these corrections. Furthermore there are no chirally enhanced power corrections in the vector meson wave functions.
%
%

\begin{theacknowledgments}
I would like to thank the organizers of 
both the Regensburg workshop and
Heavy Flavors 9 for inviting me to such interesting symposia.
It is a great pleasure to thank Gerhard Buchalla for the extremely enjoyable collaboration. For the work presented here, I also gratefully acknowledge financial support from the Studienstiftung des deutschen Volkes and thank the CERN Theory Division for the kind hospitality.
\end{theacknowledgments}

\vfill\eject

\end{document}